\documentclass[
amsmath,amssymb,
prd,nofootinbib,floatfix,12pt,
]{revtex4}

\newcommand\beq{\begin{eqnarray}}
\newcommand\eeq{\end{eqnarray}}
\def\lsim{\mathrel{\rlap{\lower4pt\hbox{$\sim$}}
    \raise1pt\hbox{$<$}}}                
\def\gsim{\mathrel{\rlap{\lower4pt\hbox{$\sim$}}
    \raise1pt\hbox{$>$}}}            

\allowdisplaybreaks
\usepackage{graphicx}
\usepackage{setspace}

\begin{document}
\renewcommand{\theequation}{\arabic{section}.\arabic{equation}}
\renewcommand{\thefigure}{\arabic{section}.\arabic{figure}}
\renewcommand{\thetable}{\arabic{section}.\arabic{table}}

\title{\large \baselineskip=20pt 
Signal-background interference for a singlet spin-0\\  digluon resonance at the LHC}

\author{Stephen P.~Martin}
\affiliation{{\it Department of Physics, Northern Illinois University, DeKalb IL 60115}}

\begin{abstract}\normalsize \baselineskip=14pt 
Dijet mass distributions can be used to search for spin-0 resonances that couple to 
two gluons. I show that there is a substantial impact on such searches from the 
interference between the resonant signal and the continuum QCD background amplitudes. 
The signal dijet mass distribution is qualitatively modified by this interference, 
compared to the naive expectation from considering only the pure resonant contribution, 
even if the total width of the resonance is minimal and very small compared to the 
experimental dijet mass resolution. The impact becomes more drastic as the total 
width of the resonance increases. These considerations are illustrated using 
examples relevant to the 750 GeV diphoton excess recently observed at the LHC.
\end{abstract}

\maketitle

\tableofcontents

\baselineskip=15.2pt

\setcounter{footnote}{1}
\setcounter{figure}{0}
\setcounter{table}{0}

\section{Introduction\label{sec:intro}}
\setcounter{equation}{0}
\setcounter{figure}{0}
\setcounter{table}{0}
\setcounter{footnote}{1}

In their LHC Run 2 data sets at $\sqrt{s}=13$ TeV, the ATLAS and CMS 
experiments have each observed \cite{digammaATLAS,digammaCMS} a local excess of 
events in $pp \rightarrow \gamma\gamma$, peaked at invariant mass near 
750 GeV. This excess cannot be explained within the Standard Model 
except by statistical fluctuation, and it has therefore provoked a very 
high momentum (massive and fast) literature seeking to interpret it. A 
recent review containing references to this literature can be found in 
ref.~\cite{Strumia:2016wys}. One obvious class of candidate models is 
reminiscent of the Standard Model Higgs boson, and consists of a new 
gauge-singlet spin-0 particle, called $X$ here, of mass 
$M \approx 750$ GeV, which couples by non-renormalizable 
dimension-5 operators to gluon pairs and to photon pairs, 
providing the production mode $gg \rightarrow X$ and the decay 
$X \rightarrow \gamma\gamma$. 

That class of models necessarily also 
predicts a dijet signal from $X \rightarrow gg$. The QCD background
for the process $pp \rightarrow jj$ is very large, and the effects of
QCD radiation, hadronization, and detector resolution are considerably less favorable
than for the diphoton signal. However, it is clear that this is a way 
to limit (or explore)
the coupling parameter space of these models. When the $Xgg$ coupling is 
too large, the model parameter point can in principle be ruled out by checking that
the dijet invariant mass spectrum is smooth and consistent with 
the predictions of the Standard Model. In this paper, I will use $X$ to refer 
more generally to a singlet spin-0 (scalar or pseudoscalar) digluon resonance, without
necessarily requiring it to be relevant to the 750 GeV diphoton excess.
 
The study of dijet mass distributions at hadron colliders has a long history,
with developments before 2012 reviewed in ref.~\cite{Harris:2011bh}. 
Because of high background rates, using the LHC to 
set limits near dijet mass $750$ GeV is somewhat more problematic than at higher masses,
and requires special handling to reduce data size 
and/or inefficiency due to trigger prescaling 
of the data sets. At this writing, the most recent limits on 750 GeV dijet resonances 
(not specialized to gauge-singlet, nor to spin 0, 
nor to decays into gluon pairs rather than quark-antiquark or quark-gluon) have been
given by ATLAS in ref.~\cite{Aad:2014aqa} and by CMS in ref.~\cite{Khachatryan:2016ecr}. 
The CMS limit, which uses a technique called data scouting to evade the
trigger pre-scaling limitations, and wide jets to improve the dijet mass resolution 
of the signal (as in a previous
CMS search \cite{Chatrchyan:2011ns}),
has been variously interpreted (see for example 
\cite{Knapen:2015dap,Franceschini:2015kwy,Gupta:2015zzs,
Falkowski:2015swt,Franceschini:2016gxv})
by theorists interested in the 750 GeV diphoton excess to imply an upper limit
of from 1 to 4 pb on the cross-section at $\sqrt{s}=8$ TeV for resonant production 
of $X$ with decays to digluons.

The purpose of the present paper is to point out that when interpreting 
such dijet searches it is important to take into account 
the interference between the resonant 
signal $gg \rightarrow X \rightarrow gg$ process and
the continuum QCD background amplitude $gg \rightarrow gg$. In general, the impact 
of signal/background interference is greatest when the continuum QCD amplitudes
are much larger than the amplitudes for the resonant production, as is the case here. 
An earlier study of QCD continuum-resonance interference for the dijet
signal in the case of a spin-1 resonance was performed in ref.~\cite{Choudhury:2011cg}.

In the analogous case of $gg \rightarrow h \rightarrow \gamma\gamma$
involving a Standard Model Higgs boson, signal-background interference 
has been studied in refs.~\cite{Dicus:1987fk,Dixon:2003yb,hint,deFlorian:2013psa,
Martin:2013ula,Dixon:2013haa,Coradeschi:2015tna,Becot:2015xzw}. 
The interference effect on the total cross-section at the leading
order (LO) was found to be very small in \cite{Dicus:1987fk}, 
but beyond the leading order it was shown \cite{Dixon:2003yb} to
suppress the diphoton 
cross-section by a few per cent for $M_h = 125$ GeV.
Apart from this effect on the total cross-section, 
it was noted in ref.~\cite{hint} that the interference causes a shift
of the invariant mass position of the Higgs diphoton peak to slightly lower mass, 
compared to the real part of the Higgs pole mass. 
It was subsequently found that in events with emission 
of a hard jet \cite{deFlorian:2013psa,Martin:2013ula} 
the shift is much smaller and goes in the opposite direction. 
Ref.~\cite{Dixon:2013haa} provided
a complete next-to-leading order (NLO) calculation, and found that the total mass shift 
is expected to be about 60 to 70 MeV, assuming Standard Model couplings.
Ref.~\cite{Dixon:2013haa} also pointed out that the diphoton mass peak shift
can, in principle, be used to place a model-independent bound\footnote{The 
far off-shell behavior of $pp \rightarrow h \rightarrow VV$  
\cite{Kauer:2012hd} can also be 
\cite{Caola:2013yja,Campbell:2013una,
Campbell:2013wga,Campbell:2014gua,Campbell:2015vwa},
and has been \cite{Khachatryan:2014iha,Aad:2015xua,Khachatryan:2016ctc}, used
to bound the Higgs width, but in a somewhat 
more model-dependent way \cite{Englert:2014aca,Englert:2014ffa,Logan:2014ppa}.}
on the width of the Higgs boson in the case that it is not Standard Model.
The mass peak shift for events with two additional jets has also been found to be small 
\cite{Coradeschi:2015tna}, so that this class of events
can provide another reference point (along
with the $ZZ^*$ 4-lepton distribution) against which the shift can be measured.
For the case of a 750 GeV resonance, the signal/background interference
in the diphoton channel has been studied in 
refs.~\cite{Jung:2015etr,Djouadi:2016ack} in the spin-0 case and
ref.~\cite{Fabbrichesi:2016jlo} in the spin-2 case, and in the $t\overline t$ final state
in ref.~\cite{Djouadi:2016ack}.
Earlier studies of signal-background interference in $t\overline t$ production
in various other new physics contexts can be found, for example, in 
refs.~\cite{ttinterferences}.
Resonance-continuum interference at hadron colliders has also been studied for
$W'$ and $Z'$ production; see 
refs.~\cite{Wprimeinterference} 
and 
refs.~\cite{Zprimeinterference} respectively for
examples.

As we will see below, the signal-background interference effect 
in $gg \to X \to gg$ is not negligible.
The interference can produce a striking qualitative as well as
quantitative difference compared to the naive Breit-Wigner $s$-channel estimate,
the more so if the total width of $X$ is larger than the partial width into two gluons,
but the effect is substantial even if $X$ decays mostly into two gluons.
This should be taken into account to correctly set limits 
(or establish a non-Standard Model contribution) using dijet mass distributions. 

\section{Signal-background interference at parton level
\label{sec:parton}}
\setcounter{equation}{0}
\setcounter{figure}{0}
\setcounter{table}{0}
\setcounter{footnote}{1}

\begin{figure}[t]
\includegraphics[width=0.75\linewidth,angle=0]{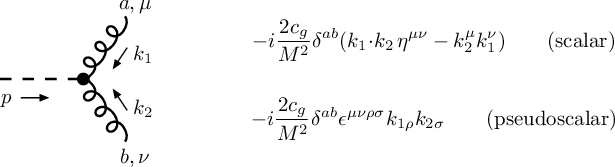}
\vspace{-0.25cm}
\caption{The Feynman rule for the effective $Xgg$ coupling, with 
$p^\mu = -k_1^\mu - k_2^\mu$, for $X$ = scalar (top) and pseudoscalar (bottom).
Here $c_g$ is a momentum-dependent form factor coupling that is approximately
constant if the interaction comes from integrating out loops of heavier particles.
\label{fig:feynmanrule}}
\end{figure}

The singlet spin-0 resonance $X$ is assumed to couple to two gluons with an 
effective field theory Feynman rule shown in Figure \ref{fig:feynmanrule},
with different expressions for the scalar and pseudoscalar cases.
In this paper, the gluons will always be on-shell and transverse. 
With this restriction, the momentum dependence of the effective form factor coupling $c_g$ 
can only be through $p^2$, the invariant mass of the digluon pair. 
If this interaction arises from loops of heavier particles,
as in the case of the Standard Model Higgs boson coupling induced by a top-quark loop, 
then in both the scalar and pseudoscalar cases for $X$ the coupling $c_g$ 
will be approximately constant. 
Because the Feynman rule scales with $p^2$, it is convenient to define
\beq
C_g(p^2) = \frac{p^2}{M^2} c_g.
\label{eq:bigClittleC}
\eeq
For simplicity, the form eq.~(\ref{eq:bigClittleC}) 
with constant $c_g$ will be assumed in the following,
although more complicated form factors could ensue 
if the particles inducing $c_g$ are not much heavier than $X$.
This coupling can then be related to the LO digluon partial width of $X$,
in both the scalar and pseudoscalar cases, by
\beq
\Gamma_{gg} = |c_g|^2/2\pi M.
\label{eq:Xwidth}
\eeq
The particle $X$ has a total width $\Gamma$, which may include other decay modes,
including $\gamma\gamma$.

At leading order, the digluon production cross-section can be written as
(including a factor of $1/2$ for identical final-state gluons):
\beq
\frac{d\sigma_{pp\rightarrow gg}}{d(\sqrt{\hat s})} &=& 
\frac{\sqrt{\hat s}}{s}
\int_{\ln\sqrt{\tau}}^{-\ln\sqrt{\tau}} dy
\, g(\sqrt{\tau} e^y) g(\sqrt{\tau} e^{-y}) 
\, 
\int_{-1}^1 dz\>\Theta(\hat s,y,z) \> \frac{d\hat\sigma}{dz}.
\phantom{xxxx}
\label{eq:dsigmadhLO}
\eeq
Here $\sqrt{s}$ is the total energy of the $pp$ collisions at the LHC,
and $\sqrt{\hat s}$ is the total partonic center-of-momentum energy, 
equal to the invariant mass of the gluon pairs in both the initial and final states. 
Also, $\tau = \hat s/s$, and $g(x)$ is the gluon parton distribution function,
$y$ is the longitudinal rapidity of the digluon center-of-momentum frame,
and $z$ is the cosine of the gluon scattering angle with respect to the proton beams. 
The factor $\Theta(\hat s,y,z)$ represents the effects of kinematic cuts. 

The naive partonic LO differential cross-section from only the $s$-channel 
resonant amplitude is, in both the scalar and pseudoscalar cases, 
\beq
\frac{d\hat \sigma_s}{dz}
&=&
\frac{|C_g(\hat s)|^4}{32 \pi \hat s D(\hat s)},
\label{eq:dsigmasdz}
\eeq
where
\beq
D(\hat s) &=& (\hat s - M^2)^2 + \Gamma^2 M^2.
\eeq
In the narrow-width approximation, one takes
\beq
1/D(\hat s) &=& \pi \delta(\hat s - M^2)/\Gamma M,
\eeq
so that after integrating over $\sqrt{\hat s}$ using the 
MSTW 2008 NLO \cite{Martin:2009iq} parton distribution function  
for the gluon with factorization scale $\mu_F = M = 750$ GeV,
one obtains a LO cross-section
\beq
\sigma (pp \rightarrow X \rightarrow gg) &\approx& 
\frac{\Gamma_{gg}^2}{M \Gamma} \times
\left \{ \begin{array}{ll}
1.06 \times 10^3\>{\rm pb} &\quad \mbox{(for $\sqrt{s} = 8$ TeV)}, 
\\
4.92 \times 10^3\>{\rm pb} &\quad \mbox{(for $\sqrt{s} = 13$ TeV).}
\end{array}
\right.
\label{eq:sigmanwa}
\eeq

However, the above is based on the narrow-width approximation without
interference effects, which is fictional. In reality, 
the complete LO partonic differential cross-section involving $X$, in excess of the
pure QCD background but including interference with it, contains other contributions:
\beq
\frac{d\hat \sigma}{dz} &=&
\frac{d\hat \sigma_s}{dz} 
+ \frac{d\hat \sigma_t}{dz} 
+ \frac{d\hat \sigma_u}{dz}
+ \frac{d\hat \sigma_{s,t}}{dz} 
+ \frac{d\hat \sigma_{s,u}}{dz} 
+ \frac{d\hat \sigma_{t,u}}{dz}
\nonumber \\ &&
+ \frac{d\hat \sigma_{s,{\rm QCD}}}{dz} + \frac{d\hat \sigma_{t,{\rm QCD}}}{dz} 
+ \frac{d\hat \sigma_{u,{\rm QCD}}}{dz} .
\eeq
The most important individual contribution in addition to eq.~(\ref{eq:dsigmasdz}) is
the interference of the $s$-channel $X$ exchange diagram with the 
pure QCD amplitude. In both the scalar and pseudoscalar cases, I find:
\beq
\frac{d\hat \sigma_{s,{\rm QCD}}}{dz}
&=&
-\frac{3 \alpha_S }{8 \hat s D(\hat s) (1 - z^2)} 
\left \{
{\rm Re}[C_g(\hat s)^2] (\hat s - M^2) + {\rm Im}[C_g(\hat s)^2] M \Gamma 
\right \}.
\label{eq:dsigmasQCDdz}
\eeq
The contribution of eq.~(\ref{eq:dsigmasQCDdz}) nearly vanishes for $\hat s = M^2$, but
has maximal excursions from 0 near $\sqrt{\hat s} = M \pm \Gamma/2$ that can be
numerically larger than the pure resonant contribution of
eq.~(\ref{eq:dsigmasdz}), especially when $\Gamma/\Gamma_{gg}$ is large. 
It has a characteristic peak-dip structure, which partially washes out
due to detector resolution and QCD radiation effects; this cancellation
is less complete when $\Gamma$ is comparable to the 
effective dijet mass resolution. More importantly,
far from the resonance mass $M$, the Breit-Wigner tails are enhanced
by the numerator factor of $\hat s - M^2$ 
in the interference term of eq.~(\ref{eq:dsigmasQCDdz}).
There is a suppression (enhancement) of the magnitude of the lower (upper) tail
from the $\hat s^2$ factor following from the momentum dependence of the $C_{g}$ coupling,
but this is counteracted by the falling 
$\hat s$ dependence of the gluon-gluon luminosity function.

The remaining contributions from diagrams with 
$t$-channel and $u$-channel exchanges of $X$, and their interferences with the
resonant $s$-channel and QCD diagrams, are numerically smaller. 
Again the scalar and
pseudoscalar cases are the same, when written in terms of $C_g$:
\beq
\frac{d\hat \sigma_{t}}{dz} + \frac{d\hat \sigma_{u}}{dz}
&=&
\frac{1}{32 \pi \hat s} \left [
\frac{|C_g(\hat t)|^4}{(\hat t - M^2)^2} + \frac{|C_g(\hat u)|^4}{(\hat u - M^2)^2}
\right ],
\label{eq:dsigmatandudzPSEUDOSCALAR}
\\
\frac{d\hat \sigma_{t,u}}{dz}
&=&
\frac{{\rm Re}[C_g(\hat t)^2 \, C_g(\hat u)^{*2}]}{256\pi \hat s
(\hat t - M^2)(\hat u - M^2)},
\label{eq:dsigmatudzPSEUDOSCALAR}
\\
\frac{d\hat \sigma_{s,t}}{dz} 
&=&
\frac{
{\rm Re}[C_g(\hat s)^2 \, C_g(\hat t)^{*2}] (\hat s - M^2) +
{\rm Im}[C_g(\hat s)^2 \, C_g(\hat t)^{*2}] \Gamma M
}{256 \pi \hat s D(\hat s) (\hat t - M^2)},\phantom{xxx}
\label{eq:dsigmastdzPSEUDOSCALAR}
\\
\frac{d\hat \sigma_{s,u}}{dz} 
&=&
\frac{
{\rm Re}[C_g(\hat s)^2 \, C_g(\hat u)^{*2}] (\hat s - M^2) +
{\rm Im}[C_g(\hat s)^2 \, C_g(\hat u)^{*2}] \Gamma M
}{256 \pi \hat s D(\hat s) (\hat u - M^2)},\phantom{xxx}
\label{eq:dsigmasudzPSEUDOSCALAR}
\\
\frac{d\hat \sigma_{t,{\rm QCD}}}{dz} 
&=&
\frac{3\alpha_S {\rm Re}[C_g(\hat t)^2] (1-z)^2}{64 \hat s
(\hat t - M^2)(1+z)} ,
\label{eq:dsigmatQCDdzPSEUDOSCALAR}
\\
\frac{d\hat \sigma_{u,{\rm QCD}}}{dz} 
&=&
\frac{3\alpha_S {\rm Re}[C_g(\hat u)^2] (1+z)^2}{64 \hat s
(\hat u - M^2)(1-z)} .
\label{eq:dsigmauQCDdzPSEUDOSCALAR}
\eeq
Above, I have neglected $i \Gamma M$ in the $t$-channel and $u$-channel propagators.
Of these contributions involving $t$-channel and $u$-channel
exchange, only the ones that involve interference with QCD 
[(\ref{eq:dsigmatQCDdzPSEUDOSCALAR})-(\ref{eq:dsigmauQCDdzPSEUDOSCALAR})]
are numerically appreciable in the examples below, but all are included for completeness. 
The large and well-known continuum pure QCD contributions to the 
differential cross-section are not shown, and in practice are modeled by the experimental
collaborations using a parameterized smoothly falling background.
In the following, I will also 
assume that there is no absorptive part of the $Xgg$ form factor,
so that $C_g(p^2)$ is always real. 

\section{Numerical results for $\sqrt{s} = 8$ TeV\label{sec:num8TeV}}
\setcounter{equation}{0}
\setcounter{figure}{0}
\setcounter{table}{0}
\setcounter{footnote}{1}

In this section, I will numerically illustrate the impact of the interference 
effects for the LHC runs with $\sqrt{s} = 8$ TeV. In order to be approximately
relevant to the boundary of the region that has sometimes been taken to be excluded
by the CMS result \cite{Khachatryan:2016ecr}, 
I will choose benchmarks that in the naive narrow-width approximation
would yield $\sigma(pp\rightarrow X \rightarrow jj) \approx 2.5$ pb at $\sqrt{s}=8$ TeV.
Using eq.~(\ref{eq:sigmanwa}) and including a $K$ factor of 1.5, this implies
$\Gamma_{gg}^2/\Gamma \approx 0.0016 M$.
Therefore, I will take as one benchmark scenario
the case that this is saturated, with $\Gamma_{\gamma\gamma}$ and all other
partial widths much smaller:
\beq
\Gamma_{gg} = \Gamma = 0.0016 M,
\label{eq:benchmark8a}
\eeq
which implies $c_g = 75$ GeV. 
If instead $X$ also has significant decay widths 
into other states, so that the total width 
$\Gamma$ is larger than $\Gamma_{gg}$,
then obtaining 2.5 pb for the narrow-width prediction cross-section requires 
larger $\Gamma_{gg}$, and therefore larger $c_g$, according to eq.~(\ref{eq:Xwidth}).
There are significant constraints on partial widths into other Standard Model 
2-body final states, including $X$ decaying to invisible
particles (see for example \cite{Franceschini:2015kwy,Franceschini:2016gxv}). 
However, if $X$ can decay into 
more nondescript (but not invisible) final states, for example collections of 
soft jets, then these constraints may not apply, and $\Gamma$ can be larger 
than $\Gamma_{gg}$. I will therefore
consider below two other benchmark cases:
\beq
\Gamma_{gg} = 0.004 M,\qquad
\Gamma = 0.01 M,
\label{eq:benchmark8b}
\eeq
and
\beq
\Gamma_{gg} = 0.01 M,\qquad
\Gamma = 0.06 M.
\label{eq:benchmark8c}
\eeq
The last case is of interest because ATLAS has found \cite{digammaATLAS} 
that the data may prefer a width to mass ratio of order 6\%.
One could even consider larger widths $\Gamma$ as being compatible with the excesses
observed by both ATLAS and CMS.

\begin{figure}[!tb]
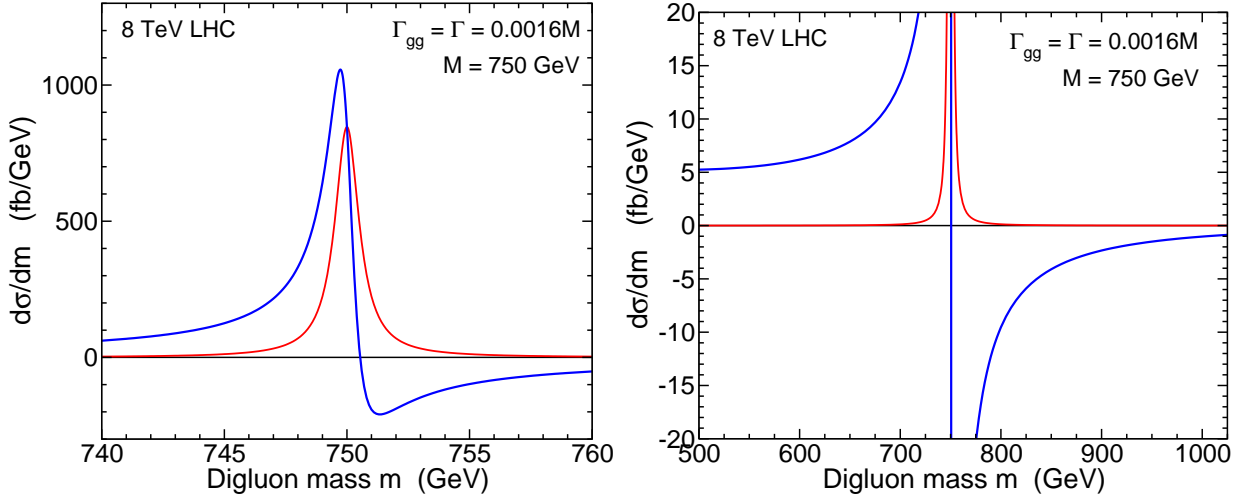

\begin{minipage}[]{0.49\linewidth}
\includegraphics[width=\linewidth,angle=0]{unsmeared_0.0016_0.0016_narrow.eps}
\end{minipage}
\begin{minipage}[]{0.49\linewidth}
\begin{flushright}
\includegraphics[width=\linewidth,angle=0]{unsmeared_0.0016_0.0016_wide.eps}
\end{flushright}
\end{minipage}
\caption{\label{fig:unsmeared0016}
The digluon invariant mass distribution for $pp\rightarrow gg$ at leading order,
for the case $\Gamma_{gg} = \Gamma_{\rm tot} = 0.0016M$.
The thinner (red) lines show the fictional result with only the 
$s$-channel resonance diagram $gg \rightarrow X \rightarrow gg$
included, while the thicker (blue) lines show the full result 
from eqs.~(\ref{eq:dsigmasdz}), (\ref{eq:dsigmasQCDdz}),
and (\ref{eq:dsigmatandudzPSEUDOSCALAR})-(\ref{eq:dsigmauQCDdzPSEUDOSCALAR})
including 
interferences with the continuum QCD $gg \rightarrow gg$ amplitude.
The two panels show the same data but with different scales on the axes.}
\end{figure}

For purposes of illustration, I 
apply parton-level cuts $p_T > 40$ GeV, and $|\eta| < 2.5$ on the gluons following
ref.~\cite{Khachatryan:2016ecr}. The resulting
digluon invariant mass distributions for $pp \rightarrow gg$,  from
the formulas in the preceding section, 
are shown in Figure \ref{fig:unsmeared0016} as a function of 
$m = \sqrt{\hat s}$. 
The thinner (red) line shows the fictional result with only the 
$s$-channel resonance diagram $gg \rightarrow X \rightarrow gg$
[from eq.~(\ref{eq:dsigmasdz})]
included, while the thicker (blue) line is the full result including 
the $t$-channel and $u$-channel exchange of $X$ and their interferences with
the continuum QCD $gg \rightarrow gg$ diagrams, from eqs.~(\ref{eq:dsigmasdz}), 
(\ref{eq:dsigmasQCDdz}),
and (\ref{eq:dsigmatandudzPSEUDOSCALAR})-(\ref{eq:dsigmauQCDdzPSEUDOSCALAR}). 
The peak of the full distribution
is somewhat larger in magnitude and slightly lower in mass than the naive resonant-only approximation, but 
the extensive positive and negative tails will be of greater importance 
after resolution effects, as shown below. Note that $d\sigma/dm$ is
negative where the interference dominates and is destructive, because
the pure continuum QCD background 
contribution, not shown, renders the total positive.
(No $K$ factors are included in these plots.)

\begin{figure}[!tb]
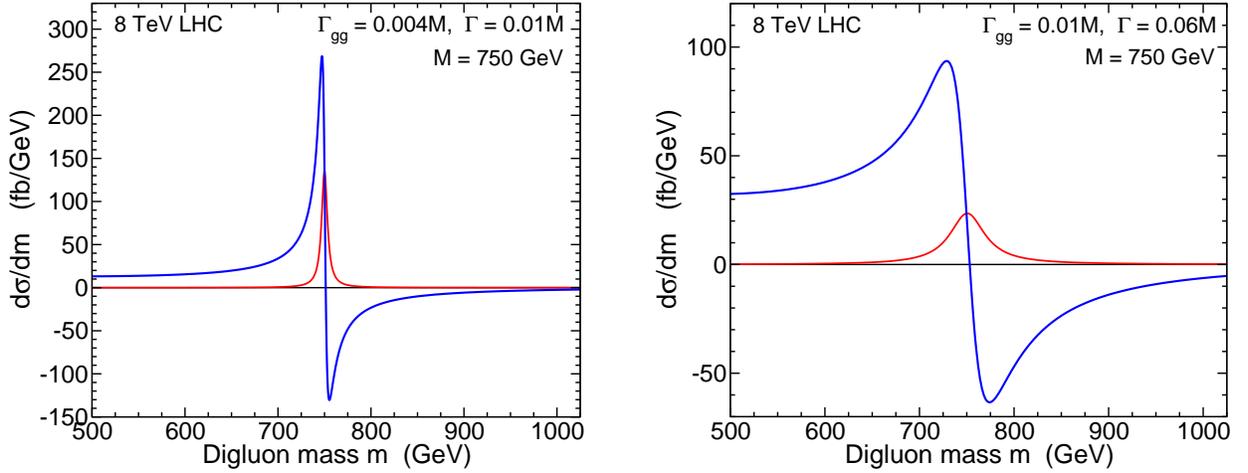

\begin{minipage}[]{0.49\linewidth}
\begin{flushleft}
\includegraphics[width=0.95\linewidth,angle=0]{unsmeared_0.004_0.01.eps}
\end{flushleft}
\end{minipage}
\begin{minipage}[]{0.49\linewidth}
\begin{flushright}
\includegraphics[width=0.95\linewidth,angle=0]{unsmeared_0.01_0.06.eps}
\end{flushright}
\end{minipage}
\caption{\label{fig:unsmearedwider} 
The digluon invariant mass distribution for $pp\rightarrow gg$ at leading order,
for the cases $\Gamma_{gg} = 0.004M$ and $\Gamma_{\rm tot} = 0.01M$ (left panel)
and $\Gamma_{gg} = 0.01M$ and $\Gamma_{\rm tot} = 0.06M$ (right panel).
The thinner (red) lines show the fictional results with only the 
$s$-channel resonance diagram $gg \rightarrow X \rightarrow gg$
included, while the thicker (blue) lines show the full results 
from eqs.~(\ref{eq:dsigmasdz}), (\ref{eq:dsigmasQCDdz}),
and (\ref{eq:dsigmatandudzPSEUDOSCALAR})-(\ref{eq:dsigmauQCDdzPSEUDOSCALAR})
including 
interference with the continuum QCD $gg \rightarrow gg$ amplitude.
} 
\end{figure}

Similarly, Figure \ref{fig:unsmearedwider} shows the parton-level 
digluon mass distributions for the cases with larger widths given in eqs.~(\ref{eq:benchmark8b}) and (\ref{eq:benchmark8c}).%
These figures show that with $\Gamma > \Gamma_{gg}$, the interference 
effect is much larger than the naive pure resonance contribution, with fatter positive and negative tails
below and above $M$. 

\begin{figure}[!t]
\begin{minipage}[]{0.4\linewidth}
\begin{flushright}
\includegraphics[width=\linewidth,angle=0]{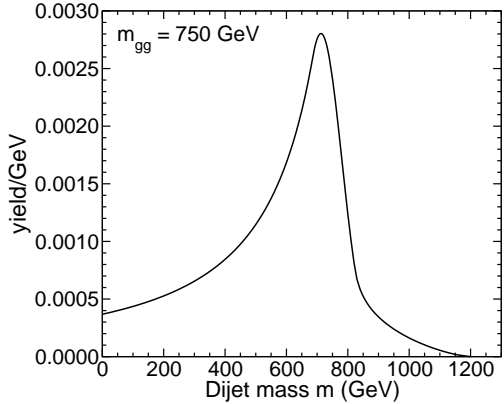}
\end{flushright}
\end{minipage}
\begin{minipage}[]{0.49\linewidth}
\begin{minipage}[]{0.04\linewidth}\end{minipage}
\begin{minipage}[]{0.90\linewidth}
\caption{\label{fig:lineshape}
The assumed normalized detector response dijet mass distribution, 
from eq.~(\ref{eq:smearfunction}), for the example case of a 
gluon pair with invariant mass $m_{gg} = 750$ GeV.}
\end{minipage}\end{minipage}
\end{figure}

In order to approximately model the detector responses for the 
dijet invariant mass distributions, below I will 
smear the final state digluon invariant 
masses by convolution with a double-sided crystal ball function 
\cite{Oreglia:1980cs}
with a cutoff at large masses, i.e., a Gaussian core smoothly
matched to power-law tails on each side. For a given input 
digluon invariant mass $m_{gg}$, 
this distribution function for the observed dijet invariant mass 
$m$ is approximated by the form:
\beq
f(m,m_{gg}) &=&
N \left \{ \begin{array}{ll}
(A_L + B_L m)^{-n_L} & \mbox{for } (m - \overline m)/\sigma \leq -\alpha_L,\\
{\rm exp}[-(m-\overline m)^2/2\sigma^2]
 \phantom{xx} &\mbox{for }  -\alpha_L \leq (m - \overline m)/\sigma \leq \alpha_H, \\
(1 - m/m_{\rm max})^\nu (A_H + B_H m)^{-n_H} 
\phantom{x} &\mbox{for }(m - \overline m)/\sigma \geq \alpha_H .
\end{array}
\right.\phantom{xxx}
\label{eq:smearfunction}
\eeq
Here, the width in the Gaussian core of the distribution is taken to be 
$\sigma/m_{gg} = 2.09/\sqrt{m_{gg}} + 0.015$ as obtained in the CMS wide 
jet analyses \cite{Khachatryan:2016ecr,Chatrchyan:2011ns}. I also use 
$\overline m = 0.95 m_{gg}$, $m_{\rm max} = 1.6 m_{\rm gg}$, 
$n_L = 1.5$, $\alpha_L = 0.4$, $n_H = 0.25$, $\alpha_H = 1.6$,
and $\nu = 1.4$, estimated to roughly match Figure 2 in 
ref.~\cite{Khachatryan:2016ecr} when $m_{gg}=900$ GeV.
The constants $A_L, B_L, A_H, B_H$ are then uniquely determined in terms of the 
other parameters by the continuity of $f(m,m_{gg})$ and its 
first derivative with respect to $m$, 
and the normalization constant $N$ is fixed by the requirement 
$\int_0^{m_{\rm max}} f(m,m_{gg})\, dm = 1$.
As an illustration, the function $f(m,m_{gg})$ is shown in 
Figure \ref{fig:lineshape} for the example case $m_{gg} = 750$ GeV.

\begin{figure}[!t]
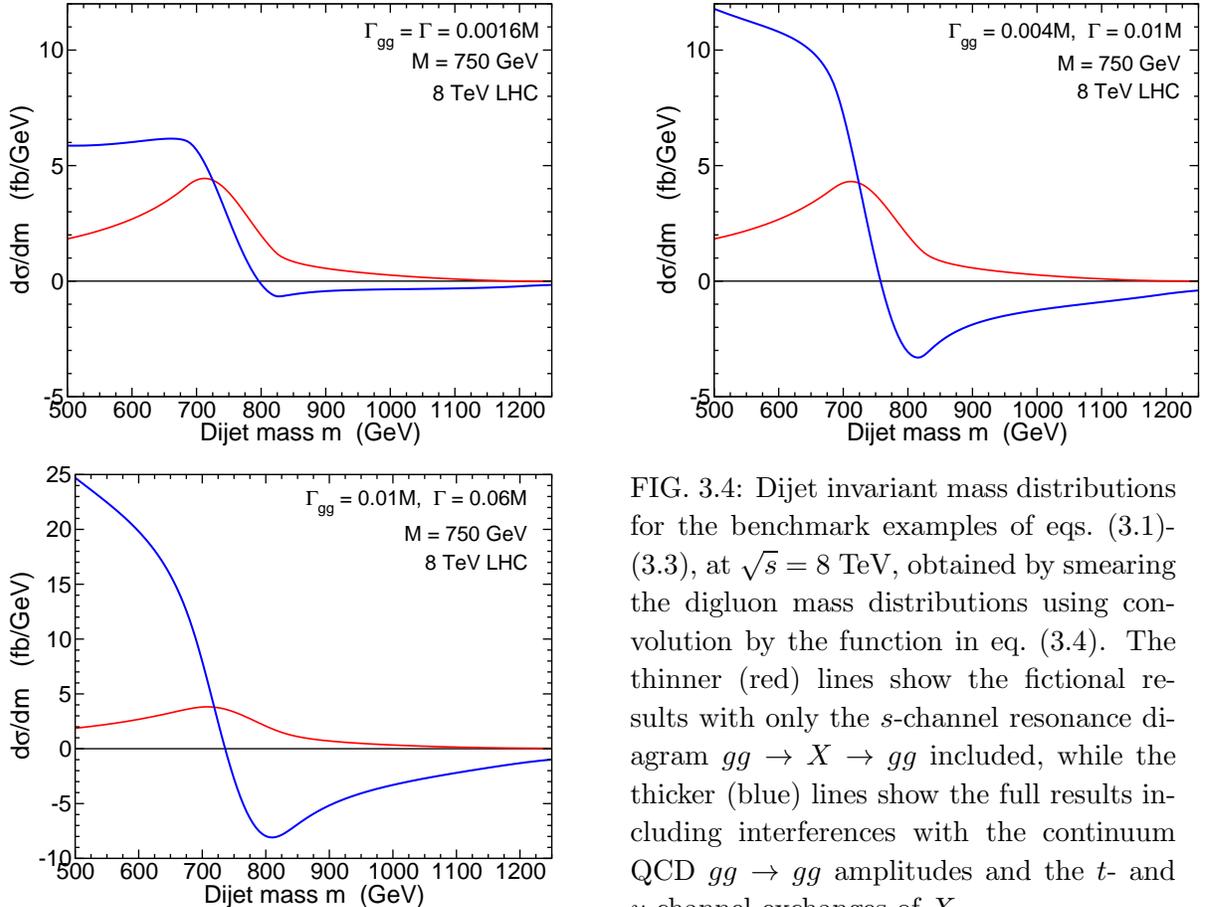

\begin{minipage}[]{0.49\linewidth}
\includegraphics[width=0.9\linewidth,angle=0]{smeared_0.0016_0.0016.eps}
\end{minipage}
\begin{minipage}[]{0.49\linewidth}
\begin{flushright}
\includegraphics[width=0.9\linewidth,angle=0]{smeared_0.004_0.01.eps}
\end{flushright}
\end{minipage}
\begin{minipage}[]{0.49\linewidth}
\includegraphics[width=0.9\linewidth,angle=0]{smeared_0.01_0.06.eps}
\end{minipage}
\begin{minipage}[]{0.49\linewidth}
\begin{minipage}[]{0.04\linewidth}\end{minipage}
\begin{minipage}[]{0.90\linewidth}
\caption{\label{fig:smeared}
Dijet invariant mass distributions for the benchmark examples 
of eqs.~(\ref{eq:benchmark8a})-(\ref{eq:benchmark8c}), at $\sqrt{s} = 8$ TeV,
obtained by smearing the digluon mass distributions 
using convolution by the function in 
eq.~(\ref{eq:smearfunction}). 
The thinner (red) lines show the fictional results with only the 
$s$-channel resonance diagram $gg \rightarrow X \rightarrow gg$
included, while the thicker (blue) lines show the full results including 
interferences with the continuum QCD $gg \rightarrow gg$ amplitudes
and the $t$- and $u$-channel exchanges of $X$.
} 
\end{minipage}\end{minipage}
\end{figure}

After smearing by convolution with 
the parton-level results using eq.~(\ref{eq:smearfunction}),
one obtains the distributions shown in Figure \ref{fig:smeared},
for the three benchmark cases described above.%
The case with $\Gamma_{gg} = \Gamma = 0.0016M$ has the smallest effect from the interference with QCD, but even in this case one sees that the resulting dijet mass distribution is very different from the naive expectation obtained by including
only the $s$-channel $X$ exchange amplitude. The maximum excursion from 0 is about 50\%
larger than the naive expectation, and is more of a plateau rather than a peak. 
In the resonance
region $m\approx M$, the full distribution falls rapidly until reaching a shallow but 
long negative tail for $m > 800$ GeV. It is not immediately 
clear how this will affect the setting of limits,
because it depends on how the QCD background is parameterized in the data analysis. 
In particular, in the case that the signal is present, the 
positive tail for $m < 700$ GeV might be absorbed into the background fit, 
leading to a smaller peak at lower $m$ and an apparent dip for $m>700$ GeV, 
rather than a pure peak. 

For the larger total width cases shown, with $\Gamma = 0.01M$ and $0.06 M$, 
the off-resonance positive and negative tails from the $\hat s - M^2$ numerator factor
in eq.~(\ref{eq:dsigmasQCDdz}) become much more pronounced, as they are enhanced 
by a larger $c_g^2$ factor. In these cases, the dip would be impossible to 
miss if the effect of $X$ is visible at all, regardless of 
the method used to model the background. 
It is clear that the interference effect is 
crucial in interpreting the dijet mass distribution in order to set search 
limits on a digluon resonance, as the naive pure resonance behavior 
is completely different from the full result, both qualitatively and quantitatively.

\section{Numerical results for $\sqrt{s} = 13$ TeV\label{sec:num13TeV}}
\setcounter{equation}{0}
\setcounter{figure}{0}
\setcounter{table}{0}
\setcounter{footnote}{1}

At this writing, the LHC is colliding protons with $\sqrt{s} = 13$ TeV, adding
to the existing data sets at that energy which gave rise to the 750 GeV diphoton excess.
Whether or not that excess is confirmed, it will be important to look 
for dijet anomalies as part of a robust program of searches for physics 
beyond the Standard Model. In this section, I consider a few benchmark cases 
for a singlet spin-0 resonance at this higher energy.

From the formulas in section \ref{sec:parton}, 
it becomes apparent that the relative shapes of the
full and naive dijet mass distributions at leading order have only a weak dependence on
$\sqrt{s}$. This is because all of the contributions to $d\sigma/dz$ 
are multiplied by the same gluon-gluon luminosity function, for a given $\sqrt{s}$. 
There is a dependence on kinematic cuts, which produce relatively minor differences 
in shape. The main dependences of the shapes of the distributions 
come instead from
the total width $\Gamma$ and the coupling strength parameterized by $\Gamma_{gg}$
according to eq.~(\ref{eq:Xwidth}).
Therefore, as benchmarks I choose four cases picked so that in each case 
the naive narrow-width approximation would give a total cross-section 
of about 3 pb from eq.~(\ref{eq:sigmanwa}) at $\sqrt{s} = 13$ TeV 
after including a $K$ factor of 1.5. Results obtained by
scaling $\Gamma$ and $\Gamma_{gg}$ by a common factor should have 
different magnitudes but roughly the same
shapes. The four chosen benchmark cases for $\sqrt{s} = 13$ TeV are:
\beq
\Gamma_{gg} &=& \Gamma \>=\> 0.0004 M,
\label{eq:benchmark13a}
\\ 
\Gamma_{gg} &=& 0.0009M,\qquad\Gamma \>=\> 0.002M,
\\ 
\Gamma_{gg} &=& 0.002M,\qquad\Gamma \>=\> 0.01M,
\\ 
\Gamma_{gg} &=& 0.005M,\qquad\Gamma \>=\> 0.06M.
\label{eq:benchmark13d}
\eeq
The first case eq.~(\ref{eq:benchmark13a})
is the minimal width for the given target 
narrow-width approximation cross-section,
while the last eq.~(\ref{eq:benchmark13d}) is included because of the present ATLAS preference for a large width
when the model is intended to explain the 750 GeV diphoton excess.

\begin{figure}[!t]
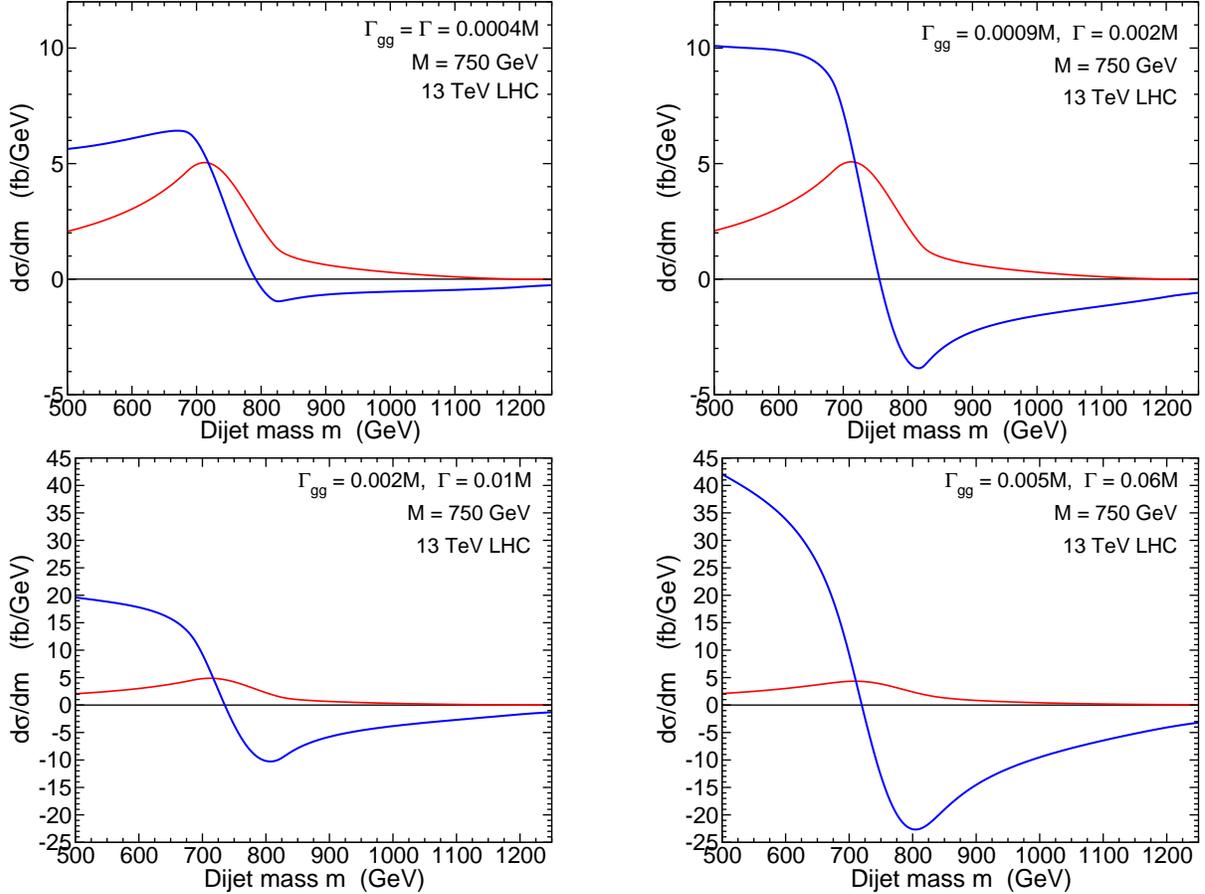

\begin{minipage}[]{0.49\linewidth}
\includegraphics[width=0.9\linewidth,angle=0]{smeared_0.0004_0.0004_13TeV.eps}
\end{minipage}
\begin{minipage}[]{0.49\linewidth}
\begin{flushright}
\includegraphics[width=0.9\linewidth,angle=0]{smeared_0.0009_0.002_13TeV.eps}
\end{flushright}
\end{minipage}
\begin{minipage}[]{0.49\linewidth}
\includegraphics[width=0.9\linewidth,angle=0]{smeared_0.002_0.01_13TeV.eps}
\end{minipage}
\begin{minipage}[]{0.49\linewidth}
\begin{flushright}
\includegraphics[width=0.9\linewidth,angle=0]{smeared_0.005_0.06_13TeV.eps}
\end{flushright}
\end{minipage}
\caption{\label{fig:smeared13TeV}
Dijet invariant mass distributions for the benchmark examples of
eqs.~(\ref{eq:benchmark13a})-(\ref{eq:benchmark13d}), 
with $\sqrt{s} = 13$ TeV,
obtained by smearing the digluon mass distributions 
using convolution by the function in 
eq.~(\ref{eq:smearfunction}). 
The thinner (red) lines show the fictional results with only the 
$s$-channel resonance diagram $gg \rightarrow X \rightarrow gg$
included, while the thicker (blue) lines show the full results including 
interferences with the continuum QCD $gg \rightarrow gg$ amplitudes
and the $t$- and $u$-channel exchanges of $X$.
} 
\end{figure}

The dijet mass distributions obtained after smearing by convolution with the
response function in eq.~(\ref{eq:smearfunction}) are shown in Figure \ref{fig:smeared13TeV}.
As promised, the minimal width case has a strong resemblance 
in shapes to the minimal width benchmark
case at $\sqrt{s} = 8$ TeV with $\Gamma= \Gamma_{gg}$
(compare the first panel of Figure \ref{fig:smeared}). This represents 
the minimal impact of the QCD interference contributions compared to the
naive $s$-channel $X$ exchange contribution. As the ratio $\Gamma/\Gamma_{gg}$ 
increases, the positive and negative tails from the QCD interference become more
dominant.

\vspace{-0.5cm}

\section{Outlook\label{sec:outlook}}
\setcounter{equation}{0}
\setcounter{figure}{0}
\setcounter{table}{0}
\setcounter{footnote}{1}

In this paper, I have argued that the effects of interference with 
the QCD continuum background must be included when searching for digluon resonances
at the LHC. The interference effects can overwhelm the naive pure resonance
contribution even if the width of the resonance is much smaller 
than the dijet mass resolution.
Although the numerical examples here were confined to the case of $M=750$ GeV,
I have checked that the results are quite similar for larger masses.

Only the leading order effects have been included here, so 
the results obtained only demonstrate the importance and the general size of the effect. 
It is clear that for a more realistic numerical estimate, it will be necessary 
to include at least 
NLO corrections with virtual 1-loop and real emission of an extra jet. In this regard, note that the real emission
contributions with an extra jet could well have a quite different structure
of interference with the QCD background, as they come in part from quark-gluon
and quark-antiquark scattering, with completely different initial and final states. 
In the somewhat analogous case of 
$gg \rightarrow h \rightarrow \gamma\gamma$ in the Standard Model, recall that 
emission of one or two additional jets (with e.g.~$p_T > 30$ GeV) results 
in a much smaller shift 
\cite{deFlorian:2013psa,Martin:2013ula,Dixon:2013haa,Coradeschi:2015tna}
in the Higgs diphoton peak compared to the LO shift
\cite{hint}. It remains to be seen how
such NLO effects behave in the present situation. It would also be interesting to 
evaluate the impact of resonance-continuum interference on dijet resonances 
with other spin
and color quantum number assignments, including for example a spin-2 singlet resonance,
or a resonance that decays to $q\overline q$.

I have not attempted a full Monte Carlo simulation of the detector responses,
which could not be as accurate as results from the ATLAS and CMS 
detector collaborations themselves.
In the search for a spin-0 digluon resonance, greater sensitivity can probably
be obtained by enforcing a harder cut on the leading two jet transverse momenta 
in addition to the generic dijet mass spectrum requirements, because the
LO pure resonance 
signal (before cuts) is isotropic in the center-of-momentum frame, while
the LO pure QCD background is forward-backward peaked, with $1/(1\pm z)^2$ 
singularities for scattering near the beam axis where $z = \mp 1$. 
Note that the LO $X$-QCD interference term is intermediate 
between these, with $1/(1\pm z)$ 
behavior as seen in eq.~(\ref{eq:dsigmasQCDdz}). 
A stringent cut on the $p_T$ for the second leading jet  
would therefore seem to be appropriate to formally maximize significance for the signal
(including interference) for a singlet spin-0 resonance over the pure QCD background
at leading order, but this should be re-evaluated for optimization
after including NLO and QCD radiation and detector resolution effects, which can 
have a strong impact on the jet $p_T$ distributions. If a dijet mass anomaly is detected 
or suspected in future data, the $p_T$ dependence 
will depend on the inclusion of interference effects, 
and could also be used to probe or limit its possible origin from a resonance.

{\it Acknowledgments:} 
I am indebted to Prudhvi Bhattiprolu for pointing out errors in 
eqs.~(\ref{eq:dsigmatandudzPSEUDOSCALAR})-(\ref{eq:dsigmauQCDdzPSEUDOSCALAR})
in the case of parity-even scalars in previous versions of this paper.
This work was supported in part by the 
National Science Foundation grant number PHY-1417028. 

\vspace{-0.25cm}
 

\end{document}